\documentclass{article}

\usepackage{amsmath,amsfonts,amssymb,times,graphicx,natbib,algorithm,algorithmic,hyperref}

\usepackage[accepted]{data4good2016}

\icmltitlerunning{Understanding Innovation}

\begin{document}

\twocolumn[
\icmltitle{Understanding Innovation to Drive Sustainable Development}

\icmlauthor{Prasanna Sattigeri}{psattig@us.ibm.com}
\icmlauthor{Aur\'elie Lozano}{aclozano@us.ibm.com}
\icmlauthor{Aleksandra Mojsilovi\'c}{aleksand@us.ibm.com}
\icmlauthor{Kush R. Varshney}{krvarshn@us.ibm.com}
\icmlauthor{Mahmoud Naghshineh}{mahmoud@us.ibm.com}
\icmladdress{IBM Thomas J. Watson Research Center,
            1101 Kitchawan Rd., Yorktown Heights, NY 10598 USA}

\vskip 0.3in
]

\begin{abstract}
Innovation is among the key factors driving a country's economic and social growth. But what are the factors that make a country innovative? How do they differ across different parts of the world and different stages of development? In this work done in collaboration with the World Economic Forum (WEF), we analyze the scores obtained through executive opinion surveys that constitute the WEF's Global Competitiveness Index in conjunction with other country-level metrics and indicators to identify actionable levers of innovation. The findings can help country leaders and organizations shape the policies to drive developmental activities and increase the capacity of innovation.
\end{abstract}

\section{Introduction}
\label{sec:intro}
Innovation plays an essential role in the development of the modern global economy. It ranks among the most important of human traits, driving economic growth through the creation of job opportunities, new products and services, motivating cities, regions and countries to create environments that foster it to improve their competitiveness in local and global markets \citep{LeadingIndicators}. Innovation is also a key component of sustainable development, and a means to uplift humanity. The United Nations (UN) has set goals for sustainable development with the aim of ending poverty, protecting the planet and bringing prosperity to all \citep{SustainableDevelopment}. UN states that one of the targets among these sustainable development goals is to “support domestic technology development, research and innovation in developing countries.” While innovation is easy to perceive, it is difficult to define, and consequently even more difficult to measure. Furthermore, it is not well understood what the economic, sociological or anthropological drivers of innovation are, and which outcomes or behaviors are results of innovative actions. 


Measuring innovation is an ongoing effort in the international community and currently there are many innovation indexes, surveys and reports. A comprehensive survey of these reports can be found in the WEF Leading Indicators of Innovation study \citep{EuropeanInnovation}. The study observes a significant diversity in how these various indexes, surveys and reports approach the topic of innovation. Some have updated their innovation index yearly to provide a basis for comparisons over time, while others have only published their innovation index once. Similarly, the geographical scope of innovation covered varies as well. The study also highlights several issues with the ongoing approaches. The first issue is that the existing indexes, surveys and reports examining innovation are predominantly focused on “input” indicators of innovation (e.g. research and development expenditures, education level of population), which measure the context, environment and enabler factors that facilitate innovation, rather than actual innovation output or performance. This heavy emphasis on input indicators limits our understanding of innovation performance; while input indicators contribute to innovation capacity, they, unlike output indicators, do not measure results of innovation. Therefore, there is a need for an approach that would allow for better quantification of the overall innovation performance or perceived level of innovation. 


This work attempts to provide a step forward towards developing an innovation index, which would enable the measurement of innovation capabilities in an ongoing, dynamic, regional and action-oriented way. We utilize a data driven approach to identify measurable drivers of innovation, based on predictive analysis between the input country level metrics, innovation indicators and perceived innovation levels. Our hope is that this work will contribute to better understanding of what makes a country innovative, that it will offer actionable guidance in improving innovation outcomes at global and country levels, and eventually lead towards the construction of an Open Innovation Index.

This paper is organized as follows. In Section \ref{sec:data} we provide an overview of datasets used and describe the indicators and innovation scores considered. In Section \ref{sec:causal} and \ref{sec:predictors}, we provide details and results of causal and predictive modeling, respectively. Conclusions and next steps can be found in Section \ref{sec:conclusion}.  

\section{Datasets}
\label{sec:data}
Our analysis seeks to discover input/output relationships between historical data on numerous country-level metric (input) and perceived levels of innovation (output). To do so, we onsider data from the Global Competitiveness Report (GCR) \citep{schwab2013global}, and country level metrics in the World Development Indicators (WDI) \citep{WDI}.

\begin{table}
	\begin{center}
		\begin{tabular}{l l}

			Macroeconomic environment &Market Size \\
			Higher education and training & Infrastructure \\
			Goods market efficiency & Health and education \\
			Labor market efficiency & Financial markets \\
			Technological readiness & Institution\\
			Business sophistication & Innovation \\
			
		\end{tabular}
	\end{center}
	\caption{The 12 pillars of the Global Competitiveness Index.}
	\label{table:gci_pillars}
\end{table}

%
%

\subsection{Global Competitiveness Report}
To capture information on the perceived level of innovation (output) we use the Global Competitiveness Report \citep{GCRs}, a yearly report published by the WEF since 2004. The report ranks countries based on the Global Competitiveness Index, (GCI), which assesses the ability of countries to provide high levels of prosperity to their citizens. This in turn depends on how productively a country uses available resources. Therefore, the Global Competitiveness Index measures the set of institutions, policies, and factors that set the sustainable current and medium-term levels of economic prosperity \citep{GCI}. Over 110 variables contribute to the index; two thirds of them come from the Executive Opinion Survey, and one third come from publicly available sources, such as the UN. This survey contains the responses of roughly 14000 business leaders from 142 economies. The GCI variables are organized into twelve pillars (see Table \ref{table:gci_pillars}), with each pillar representing an area considered as an important determinant of competitiveness. Each of the pillars is further divided into several sub-components, which help measure that pillar. Of particular interest to this analysis is the 12th pillar: Innovation. We will use this pillar score as the ground truth for a country's innovation score.

		


\subsection{World Development Indicators}
World Development Indicators form the primary World Bank collection of development indicators, compiled from officially recognized international sources. It presents the most current and accurate global development data available, and includes national, regional and global estimates. This statistical reference includes over 1500 indicators covering more than 150 economies. The annual publication is released in April of each year, and the online database is updated three times a year. The World Bank’s Open Data site provides access to the WDI database free of charge to all users. A selection of the WDI data is featured at data.worldbank.org. We will use the statistics provided by these indicators as inputs to our analyses.





\section{Causal Analysis}
\label{sec:causal}
\begin{figure*}[!htb]
	\centering
	\includegraphics[width=\linewidth]{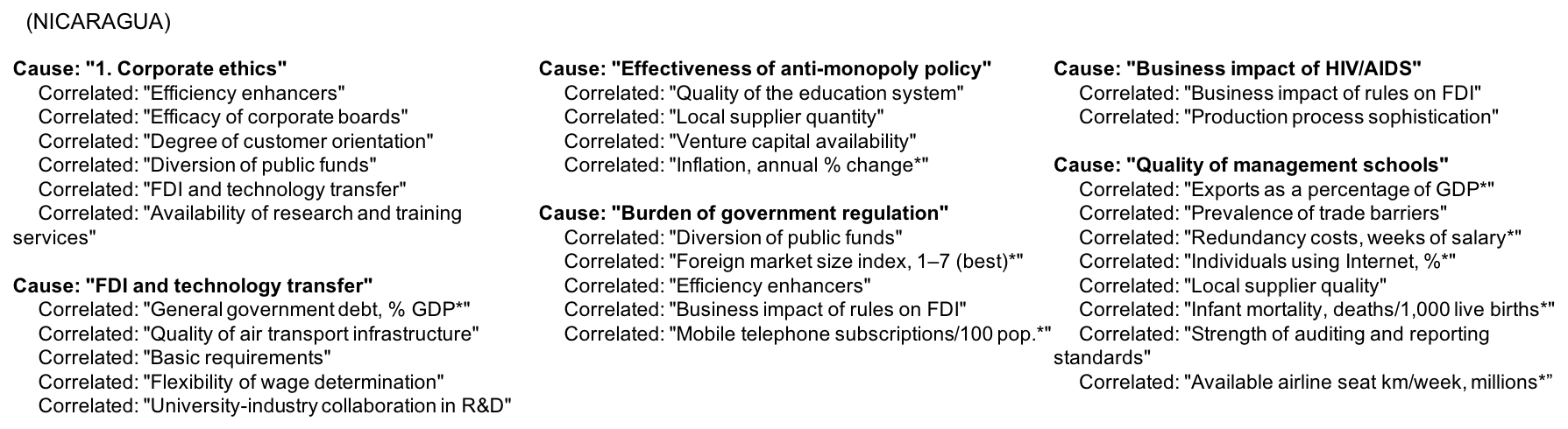}
	\caption{Top non-innovation GCI metrics for Nicaragua with  causal relationship to the overall innovation score.}
	\label{fig:causes}
\end{figure*}

To identify indicators that are causally related with innovation measurements, we perform an analysis based on  Granger causality \citep{granger1988some}.  Granger causality is a notion studied in the statistics, econometrics, machine learning and data mining literatures. The method utilizes time series to understand which factors affect other factors in the future. The main argument is that if a time series $a$ significantly helps improve the prediction of the future values of time series $b$, then $a$ is a potential cause of $b$.

The country level analysis considers a single country, and attempts to identify factors that are causally related to its innovation output. The country level analysis will shed light on a lever a particular country might be applying (either favorably or adversely) that is causally related to its level of innovation. Note, however, that if for a given country there is no activity in a particular metric (for example, no changes in R \& D investment, or growth in Internet users), despite its potential relevance to innovation, this metric will not be identified as causal indicator for the given country. In our analysis, we test if the past WDI metrics are predictive of the future innovation score. To do so, we use the time series data from 2007-2014. 

Let us consider a country $C$ with $N$ WDI metrics. Each metric is represented as a time-series with $T$ time-points. Each country also has a time-series $S$, which corresponds to the innovation score. At a time $t$, we would like to predict the innovation metric value using the past $d$ values $t - 1$ to $t - d$ of the innovation metric and the past $d$ values of the WDI metrics. We also perform the same analysis by using the non-innovation related GCI metrics. The prediction problem is solved using as a sparse linear regression approach described by \citet{sindhwani2010block}. Sparse linear regression approaches jointly perform variable selection and parameter estimation. The coefficients for the selected variables are indicative of the causal effect on the innovation score. We can further exploit the group structure among the lagged temporal variables, which is imposed by the time series they belong to. For example, lagged variables $x(t-1), x(t-2), x(t-3)$, etc., of the same time series $\{x(t)\}$ can be considered to form a group of related variables, given that they are derived from the same metric. Leveraging the group structure information leads to a more faithful implementation of the Granger causality test, as we would look at the complete time-series of a metric to determine its coefficient, instead of just looking at a lagged value of a metric. The group structure is thus imposed in terms of variable selection. If a variable is selected, then all the lagged values of this variable are also selected, and vice versa. 


In our analysis, the stopping point for group selection is tuned using approximate $C_p$ criterion \citep{yuan2006model}  to maximize the regression performance. We found that in this work, the lag value $d = 3$ provided optimal results considering the trade-off in sample size, the structure of the WDI and GCI data, and availability of metrics. The examples of causal relationships between the non-innovation GCI metrics and overall innovation score for Nicaragua  are given in Figure \ref{fig:causes}. Note that in the case of correlated features, the Granger causality test will typically select one (or few) features from the group. Given that the objective of this work is to identify all levers of innovation, for each causal factor discovered we also include up to ten metrics that are strongly correlated with it. The correlations are computed at the country level, hence depending on a county, the same metric can have different correlates.

\section{Predictors of Innovation}
\label{sec:predictors}
\begin{figure*}[!htb]
	\centering
	\includegraphics[width=0.9  \linewidth]{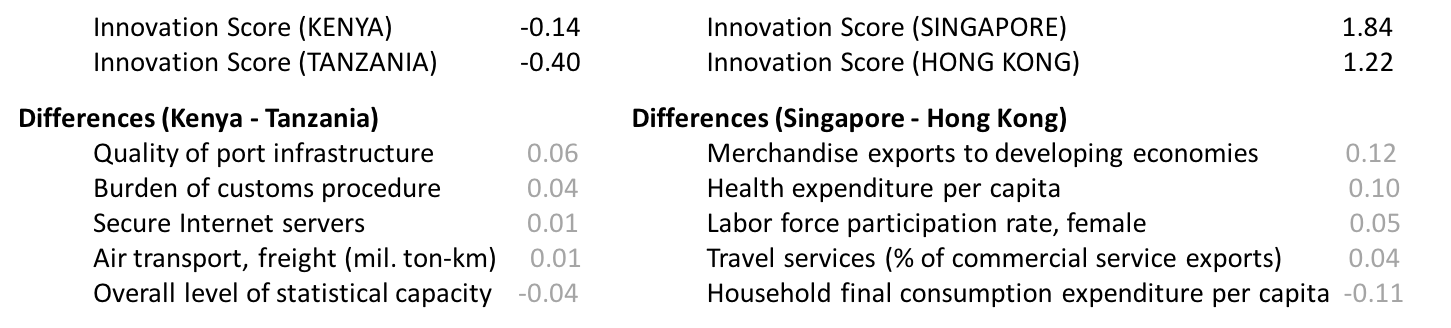}
	\caption{Innovation score prediction using random forest for a pair of countries and variables with largest differing contributions for prediction.}
 \label{fig:rf_path}
\end{figure*}

The most reliable measurements of innovation today are collected manually via executive surveys such as GCI. In this section we consider the problem of constructing an innovation index, a measurement that can be formed automatically from easily collectible metrics, thereby allowing us to benchmark countries without the burden of conducting opinion surveys. In order to enhance our understanding of how various indicators translate into the innovation scores, we formulate an appropriate prediction problem. We consider the overall innovation score from the GCI as our target variable and try to predict it using the country level indicators from WDI and non-innovation related GCI metrics. The predictive model can tell us the relative importance of different indicators in regards to the innovation score, it can help develop intuition on what factors are important for innovation, and finally, aid in identifying the metrics that should contribute to the Open Innovation Index.

To predict the innovation score, we use the Random Forest (RF) algorithm \citep{breiman2001random}. For all countries in our data set we perform prediction of the GCI innovation score using the WDI indicators. Furthermore, we also consider the prediction of the GCI innovation score using the GCI non-innovation metrics. Although not necessarily important for constructing the index, the later analysis is useful in further understanding innovation and its levers. A total of 462 WDI metrics and 162 GCI metrics that are consistently available through years 2007 to 2014 are used for the predictive modeling. Each metric is standardized to have zero mean and unit variance. Missing values are filled with 0. In the case of GCI metrics, we achieve an $R^{2}$ value of 0.93. In the case of WDI metrics, we achieve an $R^{2}$ value of 0.88. 



\begin{table}[!htb]
	\begin{center}
		\begin{tabular}{l l}
			
			\hline
			\textbf{Cluster} & \textbf{Countries} \\ \hline
			Cluster 1 & Algeria, Angola, Burundi, Gabon, Guinea, \\
			               & Haiti, Libya, Myanmar, Timor-Leste, Yemen \\
			Cluster 2 & Australia, Austria, Belgium, Canada, France, \\
			               & Iceland, Ireland, Luxembourg, Norway, \\
			               & Switzerland, United Kingdom, United States \\
			Cluster 3 & Bulgaria, Chile, Colombia, Costa Rica, \\
						  & Croatia, Ecuador, Lithuania, Panama, Peru \\
			Cluster 4 & Brunei Darussalam, Kuwait, Saudi Arabia \\
			Cluster 5 & Brazil, China, Hungary, India, Indonesia, \\
			                & Malaysia, Mexico, Poland, Russia, \\
			                & South Africa, Thailand, Turkey, Vietnam \\ \hline
			
		\end{tabular}
	\end{center}
	\caption{Constituent countries of a few clusters obtained using innovation contribution representation of the countries. }
	\label{table:country_group}
\end{table}

\subsection{Contribution Analysis}
We can further probe into the RF model learned and look at the decision path for a particular country in predicting its innovation score. This can help us understand which metrics were crucial in predicting the innovation score for a particular country and allow us to do comparative analysis between a pair of countries, or among groups of countries. The decision path can be described in terms of the contribution made by each metric towards its innovation score. Let the number of metrics be $K$ and $c^{x}_{i}$ denote the contribution of the $i^{th}$ metric towards the innovation score $y^{x}$ for a country $x$. Then the innovation score can be obtained as the sum of all the $c^{x}_{i}$. Note that the contribution value of a variable, unlike in a linear model, is not global and depends on other variables and is specific to a particular data sample. 

The representation of a country $x$ using the contributions towards innovation score denoted as $[c^{x}_{1} ,.., c^{x}_{K}]$. This representation disentangles each WDI factors influence over innovation and hence is highly informative and meaningful. We use this representation to cluster countries into group of countries which are not only at the same innovation level but also have similar mechanisms in play while reaching that level. The k-means algorithm \citep{arthur2007k} is employed with the number of clusters set to $20$. Table \ref{table:country_group} shows the constituent countries of a few clusters. These groups of countries can then be used to compare a country within its group or across other groups to get a more insightful comparative analysis in terms of innovation drivers.

Figure \ref{fig:rf_path} shows the WDI metrics with large differences in contribution values for pairs of similar countries. The first pair we consider is Kenya and Tanzania, and the second is Singapore and Hong Kong. In each pair, the constituent countries are similar in terms of WDI metrics and have relatively close predicted innovation scores. In the case of Kenya and Tanzania, quality of port infrastructure and burden of customs procedure have more positive impact on the prediction of the innovation score for Kenya than for Tanzania. The similar can be said for merchandise exports metric of Singapore in comparison to Hong Kong. The large positive effects of other metrics are balanced by household final consumption expenditure per capita metric in favor of Hong Kong. Such analysis can help decision makers obtain more targeted insights where they want to compare a country with a similar country based on a benchmark or a country in the same geographical location or similar development stage.

\section{Conclusions}
\label{sec:conclusion}

In this work, we proposed a set of analyses that would contribute to better understanding of innovation, how to measure it, and how to drive actionable insights for diverse countries and diverse innovation conditions around the world. Our approach is data-driven, and aims to produce innovation measurements that are repeatable, systematic and objective, and can lead to dynamic country-level benchmarks. This would allow policy makers and organizations like WEF to more efficiently shape policy and interventions in under-developed countries in order to increase their developmental activity and capacity for innovation. 

As future work, we plan enrich our dataset by incorporating more indicators and longer history and also enhance our causal and predictive models to explicitly handling correlated indicators. This will improve the robustness of the models and make them more reliable. We also plan to build visualizations that would allow us to communicate actionable insights and country-level benchmark, in an easily consumable way.

\bibliography{refs}
\bibliographystyle{icml2016}

\end{document}